\documentclass{aa}
\usepackage{graphicx}
\usepackage{natbib}
\usepackage{amsmath}
\usepackage{amssymb}
\usepackage{amsfonts}
\bibpunct{(}{)}{;}{a}{}{,}

\newcommand{\Om}{{\cal{O}}}

\newcommand{\ab}{\boldsymbol{a}}

\newcommand{\xb}{\boldsymbol{x}}

\newcommand{\Kb}{\boldsymbol{K}}

\newcommand{\wb}{\boldsymbol{w}}

\newcommand{\xdb}{\boldsymbol{\dot x}}

\newcommand{\betab}{\boldsymbol{\beta}}

\newcommand{\Sigmab}{\boldsymbol{\Sigma}}

\begin{document}

   \title{Stochastic modeling of kHz QPO light curves}

   \author{R. Vio\inst{1},
           P. Rebusco\inst{2},
           P. Andreani\inst{3},
          H. Madsen\inst{4}
    \and
           R.V. Overgaard\inst{5}
          }
   \institute{Chip Computers Consulting s.r.l., Viale Don L.~Sturzo 82,
              S.Liberale di Marcon, 30020 Venice, Italy\\
              \email{robertovio@tin.it},
         \and
	      Max Planck Institut f\"ur Astrophysik, K. Schwarzschild str. 1,
	      D-85748 Garching b. M\"unchen, Germany\\
	      \email{pao@mpa-garching.mpg.de}
         \and
             INAF-Osservatorio Astronomico di Trieste
             via Tiepolo 11, 34131 Trieste, Italy \\
              \email{andreani@ts.astro.it}
         \and
             Department of Informatics and Mathematical Modelling, Technical University of Denmark, 
             Richard Petersens Plads, DK-2800 Kgs. Lyngby, Denmark \\
             \email{hm@imm.dtu.dk}
         \and
             Department of Informatics and Mathematical Modelling, Technical University of Denmark, 
             Richard Petersens Plads, DK-2800 Kgs. Lyngby, Denmark \\
             \email{rvo@imm.dtu.dk}
             }

\date{Received .............; accepted ................}

\abstract{The Klu\'zniak \& Abramowicz model explains              
high frequency, double peak, "3:2" QPOs observed in
neutron star and black hole sources in terms of a
non-linear parametric resonance between radial and
vertical epicyclic oscillations of an almost Keplerian
accretion disk. The $3:2$ ratio of epicyclic frequencies
occurs only in strong gravity. \citet{reb04} and \citet{hor04}
studied the model analytically: they proved that a
small forcing may indeed excite the parametric 3:2       
resonance, but they have not explained the physical
nature of the forcing. Here we integrate their equations
numerically, dropping the ad hoc forcing, and adding
instead a stochastic term to mimic the action 
of the very complex processes that occur in disks as, for example,
 MRI turbulence. We
demonstrate that the presence of the stochastic term
triggers the resonance in epicyclic oscillations of nearly
Keplerian disks, and influences their pattern.    
}
\titlerunning{Stochastic modelling of QPOs}
\authorrunning{R. Vio et al.}
\maketitle

\section{Introduction} \label{sec:introduction}

Quasi Periodic Oscillations (QPOs) are a common phenomenon in nature.
In the last few years many kHz QPOs have been detected in the
light curves of about $20$ neutron star and few black hole sources \citep[for a recent review, see ][]{van04}. 
The nature of these QPOs is one of the mysteries which still
puzzle and intrigue astrophysicists: apart from giving important
insights into the disk structure and the mass and spin of the central
object \citep[e.g. ][]{abr01,asc04,tor05}, they offer an
unprecedented  
chance to test Einstein's theory of General Relativity in strong fields.

High frequency QPOs lie in the range of orbital frequencies of geodesics just few Schwarzschild radii 
outside the central source. This feature inspired several models based directly on   
orbital motion \citep[e.g. ][]{ste98, lam03}, but there are also models that are based on 
accretion disk oscillations \citep{wag01, kat01,rez03, lin04}. The  
Klu\'zniak \& Abramowicz resonance model \citep[see a collection 
of review articles in ][]{abr05} stresses the
importance of the observed 3:2 ratio, pointing out that the
commensurability of frequencies is a clear signature of a 
resonance. The relevance of the $3:2$ ratio and its intimate bond with the QPOs fundamental nature is
supported also by recent observations: Jeroen Homan of MIT reported at the AAS meeting
on the 9th of January 2006 that the black hole candidate GRO $J1655-40$ showed in 2005
the same QPOs (at $\sim 300$ Hz and $\sim 450$ Hz) first detected by \citet{str01}.

The main limitation of the resonance model is that it
does not yet explain the nature of the physical mechanism
that excites the resonance. The idea that turbulence                          
excites the resonance and feeds energy into it                  
\citep[e.g. ][]{abr05} is the most natural one, but it has
never been explored in detail.
The turbulence in accretion disks is most probably due to
the  Magneto-Rotational Instability \citep[MRI; ][]{bal91}.
 At present, numerical simulations of      
turbulence in accretion disks do not fully control all   
the physics near the central source. For this reason,    
they cannot yet address the question of whether MRI
turbulence does play a role in exciting and feeding the 
$3:2$ parametric resonance.
  A situation like
this is not specific of astronomy, but it is shared by other fields in applied research and engineering.
The most common and, at the same time, effective, solution consists of modelling the unknown processes as stochastic 
ones. Such processes are characterized by a huge number of degrees of freedom and therefore they
 can be assumed to have a stochastic nature \citep[e.g. ][]{gar99}. Lacking any {\it a priori} knowledge,
the most natural choice is represented by Gaussian white-noise processes. Of course, such an assumption
is only an approximation. However, it can provide an idea of the consequences on the 
system of interest of the action of a large number of complex processes. This approach leads to the modelling
of physical systems by means of {\it stochastic differential equations} (SDE) 
\citep{may79, may82, gha91, gar99, vio05}.

The present paper is a first qualitative step in this direction in the context of QPO modelling. In
Sec.~\ref{sec:model} we synthesize a stochastic version of the non-linear resonance model. 
Some experiments are presented and discussed in Sec.~\ref{sec:discussion}. The last section summarizes our findings.
Since SDEs are not
yet very well known in astronomy, Appendix~\ref{sec:SDE} provides a brief
description of the techniques for the numerical integration that are relevant for practical applications. 

In all the experiments, we adopt the units $r_G=2GM/c^2=1$ and $c=1$.
  
\section{A simplified model for kHz QPOs} \label{sec:model}
\subsection{The Klu\'zniak-Abramowicz idea} \label{subsec:resonance}
The key point of the mechanism proposed by \citet{abr01} is the observation
that kHz QPOs often occur in pairs, and that the centroid frequencies of these
pairs are in a rational ratio \citep[e.g., ][]{str01}. This feature suggested to them that
high frequency QPOs are a phenomenon due to non-linear resonance. The analogy of radial
and vertical fluctuations in a Shakura-Sunyaev disk with the Mathieu equation pointed out
that the smallest (and hence strongest) possible resonance is the $3:2$.
In all four micro-quasars which exhibit double peaks, the ratio of the two frequencies is $3:2$,
as well as in many neutron star sources. Moreover, combinations of frequencies and sub-harmonics 
have been detected: these are all signatures of non-linear resonance. A confirmation of the fact that
kHz QPOs are due to orbital oscillations comes from the scaling of the frequencies with $1/M$, where $M$
is the mass of the central object \citep{mcc04}.
        
\subsection{Dynamics of a test particle} \label{subsec:test}

A simple mathematical approach to this idea  was first developed by \citet{reb04} and 
\citet{hor04}, in the context of isolated test particle dynamics.

The time evolution of  perturbed nearly Keplerian geodesics  is given by
\begin{align}
\ddot z(t) + \omega_{\theta}^2 z(t) & = f[\rho(t), z(t), r_0,\theta_0]; \label{eq:eq1}\\
\ddot \rho(t) + \omega_r^2 \rho(t) & = g[\rho(t), z(t), r_0, \theta_0]. \label{eq:eq2}
\end{align}
Here $\rho(t)$ and $z(t)$ denote small deviations from the circular orbit $r_0,\theta_0$
(radial and the vertical coordinates  respectively), $f$ and
 $g$ account for the coupling, and $\omega_{\theta}$ and $\omega_r$ are the epicyclic frequencies.
In the case of the Schwarzschild metric, a Taylor expansion to third order leads to:
\begin{align}
f(\rho, z, r_0, & \theta_0) \nonumber \\
& = c_{11} z \rho + c_b \dot z \dot \rho + c_{21} \rho^2 z + c_{1b} \rho \dot z
\dot \rho + c_{03} z^3; \label{eq:eq3}\\
g(\rho, z, r_0, & \theta_0) = e_{02} z^2 + e_{20} \rho^2 + e_{z2} \dot z^2 + e_{30} \rho^3 \nonumber \\
& + e_{1ze2} \rho \dot z^2
+ e_{12} \rho z^2 + e_{r2} \dot \rho^2 + e_{1re2} \dot \rho^2 \rho \label{eq:eq4}.
\end{align}
The functional form of the coefficients $c_i$ and $e_j$ can be found in \citet{reb04}.
They are constants, which depend on $r_0$, the distance of the unperturbed orbit from the
centre. In previous studies these non-linear differential equations have
been integrated numerically \citep{abr03} and analyzed through a perturbative method. 
These coupled harmonic oscillators display internal non-linear resonance, the
strongest one occurs when $ \omega_\theta:\omega_r=3:2$ and the observed
frequencies are close (but not equal) to the epicyclic ones.

\subsection{Additional terms}

As we have seen the perturbation of geodesics opens up the possibility
of internal resonances.
However these epicyclic oscillations would not be detectable without any source of energy
to make their amplitudes grow. In \citet{abr03} and \citet{reb04} this source of energy was inserted
by introducing a parameter $\alpha$. The effect of forcing (e.g., due to the
neutron star spin), and its potential to produce new (external) resonances, have been addressed recently 
(e.g. Abramowicz 2005). 
The main limit in the approach proposed by \citet{abr03} and \citet{reb04} 
is that it represents an {\it ad hoc} solution. Moreover, as stressed in Sec.~\ref{sec:introduction},
it does not consider the many processes that  take place in the central region of an accretion disk as, 
for example, MRI-driven turbulence \citep{bal91}. For this reason,
we propose the {\it stochasticized} version of Eqs.~\eqref{eq:eq1}-\eqref{eq:eq2} 
\begin{align}
\ddot z(t) + \omega_{\theta}^2 z(t) - f[\rho(t), z(t), r_0, \theta_0] & = \sigma_z \beta; \label{eq:eq1p}\\
\ddot \rho(t) + \omega_r^2 \rho(t) - g[\rho(t), z(t), r_0, \theta_0] & = 0, \label{eq:eq2p}
\end{align}
with $\sigma_z$ a constant and $\beta(t)$ a continuous, zero mean, unit variance, Gaussian white-noise process.

There is no full understanding of turbulence in accretion disks. We know that the radial component is fundamental 
in producing the effective viscosity which allows accretion to occur, and
that MRI-turbulence should be different in the vertical and radial direction. 
Here we make a first step by introducing a noise term only along the
vertical direction: in the end this ansatz alone gives interesting
results. 
In the Shakura \& Sunyaev model \citep{ss73} the turbulent viscosity 
is parametrized via the famous $\alpha_{SS}$. It is reasonable to assume that  $\sigma_z$ is at maximum
a fraction, smaller than $\alpha_{SS}$, of the disk height. Hence for a geometrically
thin disk one would expect a maximum $\sigma_z$  $\sim 10^{-4}$--$10^{-3}$.
As shown in Appendix~\ref{sec:SDE}, the smallness of the stochastic perturbation permits the development of efficient 
integration schemes for the numerical integration of the system~\eqref{eq:eq1p}-\eqref{eq:eq2p}.

\section{Results}\label{sec:discussion}

We explored the dynamics of the test particle for different values of $\sigma_z$ and initial conditions 
$z(0)$ and $\rho(0)$.
All the integrations are performed by means of the scheme~\eqref{eq:rk0}-\eqref{eq:rk4}, with
$h = 5 \times 10^{-4}$ and $t=10^5$, for $r_0=27/5$ which is the value
for which the unperturbed frequencies are in a $3:2$ ratio.  
\begin{table} 
\begin{center}
\begin{tabular}{c |c c c}	
\hline
\hline
{}	    & $0$          & $10^{-5}$ & $10^{-4}$ \\
\hline
$1$         & /            & /         & $\sim \omega_{\theta}^*/3$   \\
$2$   & $\sim 3 \omega_r^*$ & $\sim \omega_\theta^*/3$,$\sim 2 \omega_\theta^*$ & $\sim \omega_\theta^*/3$  \\	
$3$ & $\sim \omega_\theta^*/3$,$\sim 2 \omega_\theta^*$  & $\sim \omega_\theta^*/3$,$\sim 2 \omega_\theta^*$ & $\sim \omega_\theta^*/3$       \\
\hline
\end{tabular}
\end{center}
\caption{Resonant frequencies (apart from the epicyclic ones) for different initial conditions ($1,2,3$) and
 noise standard deviation ($\sigma_z=0,10^{-5}, 10^{-4}$). }
\label{tab_results}
\end{table}
As a sample, three different
starting values $[z(0),\dot z(0), \rho(0), \dot \rho(0)]$ have been
used: $[0.01, 0, 0.01, 0]$, $[0.1, 0, 0.1, 0]$, and
$[0.2, 0, 0.2, 0]$, which we refer to as models $1$, $2$ and $3$ respectively. 
For each of them we considered three values of $\sigma_z$: $0$, $10^{-5}$ and $10^{-4}$.
As pointed out in the previous section, a noise level stronger than these is unlike to occur, since it would
destabilize the accretion flow.
Moreover when the initial perturbations $z(0)$ and $\rho(0)$ are  greater than about $0.5$ they
diverge, even in absence of noise: this is the limit for which the system can
be considered weakly non-linear and physically meaningful.

\begin{figure}
        \resizebox{\hsize}{!}{\includegraphics{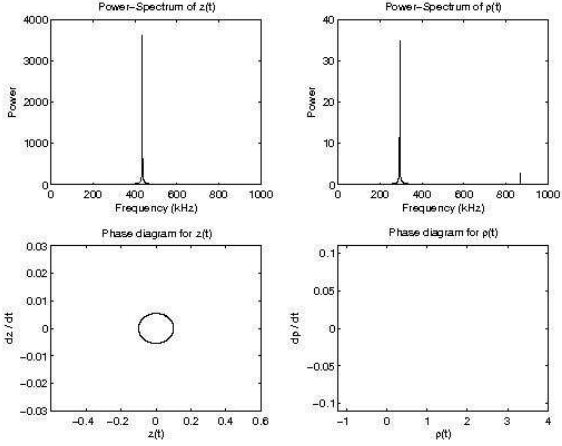}}
        \caption{Numerical simulation of the system~\eqref{eq:eq1p}-\eqref{eq:eq2p}. The upper panels show
        the power spectra of $z(t)$ and $\rho(t)$, whereas the lower panels show the 
        corresponding phase diagrams. Here $\sigma_z = 0$ (i.e. noise-free system). The displacements are in units of 
        $r_g$, the frequencies are scaled to kHz (e.g. assuming a central mass $M$ of $2 M_\odot$).}
        \label{fig:nonoise}
\end{figure}

\begin{figure}
        \resizebox{\hsize}{!}{\includegraphics{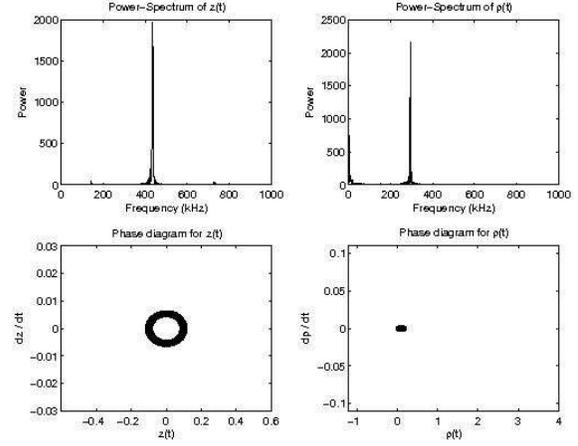}}
        \caption{The same as in Fig.~\ref{fig:nonoise} but with $\sigma_z=10^{-5}$.}
        \label{fig:noise1}
\end{figure}

\begin{figure}
        \resizebox{\hsize}{!}{\includegraphics{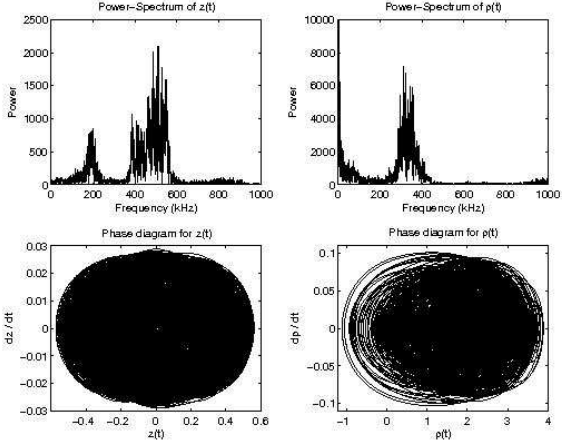}}
        \caption{The same as in Fig.~\ref{fig:nonoise} but
        $\sigma_z=10^{-4}$. The comparison with the  phase
        diagrams in the previous plots indicates how much the
        amplitudes grow under the effect of slightly stronger noise.}
        \label{fig:noise2}
\end{figure}

The lower panels of Figs.~\ref{fig:nonoise} - \ref{fig:noise2} show how the amplitudes
reach greater values for greater noise dispersion. These plots are done for the initial
conditions $2$, but similar behavior is also obtained for different initial conditions: as expected, 
noise triggers the resonances.
With regard to the frequencies at which the resonances are excited,
the dominant one are always the
epicyclic frequencies (the strongest peaks in the upper part of the
plots). However, the sub- and 
super-harmonics also react (see Tab.~$1$), and their signal is stronger for greater noise dispersion.
As predicted by means of the perturbative method of multiple scales, the dominant oscillations 
have frequencies ($\omega_r^*$ and $\omega_\theta^*$), close to the
epicyclic ones. The pattern of the other resonances (Tab.~$1$) is not
interesting in itself, as it depends on the initial conditions and on
the noise, but it is significant from a qualitative point of view, as it is a 
 signature of the non-linear nature of the system.

When the noise is $\sim 10^{-3}$ or greater the solution diverges, whilst
when it is too small ($\sim 10^{-6}$) it does not differ too much from the results
without noise. The exact limit of $\sigma_z$ over which the epicycles are swamped
depends on the initial conditions: it is indeed lower for greater initial conditions, and
vice versa.

In the case where noise is assumed to be due to MRI turbulence, this simple experiment constrains its 
amplitude: turbulence that is too low does not supply enough 
energy to the growing resonant modes, whilst too much turbulence
prevents the quasi-periodic behavior from occurring.
From this oversimplified model we get an indication that the standard deviation of vertical
MRI must be $\sim 10^{-5}-10^{-4}$, which is reasonable since it is comparable with
a small fraction of the disk height.

In a yet unpublished work
\citep[private communication,][]{ski05} considers how far
the data from a QPO source can constrain the properties of a simple damped
harmonic
oscillator model - not only its resonant frequency and damping
but also to some
extent the excitation. Not unlike the present work, he adds
random delta function shots to a simple harmonic oscillator equation,
changing the amplitude and frequency of shots. He observes that the data
constrain the allowed range of parameters for the excitation.

\section{Conclusions}\label{sec:concl}

Up to now models for kHz QPOs have been based on deterministic differential equations. The main
limits of these models is that they correspond to unrealistic physical scenarios where the many and complex 
processes that take place in the central regions of an accretion disk are not taken into account.
In this paper, we have partially overcome this problem by adopting an approach based on stochastic differential
equations. The assumption is that the above mentioned processes are characterized by a huge number of degrees of
freedom, hence they can be assumed to have a stochastic nature. In particular, we have 
investigated a simplified model for the Klu\'zniak-Abramowicz non-linear theory and shown that 
a small amount of noise in the vertical direction can trigger coupled epicyclic oscillations.
On the other hand too much noise would disrupt the quasi-periodic motion.
This is similar to the stochastically excited p-modes in the Sun \citep{gk77}.

From our simple example we get an indication that the standard deviation of vertical
noise cannot be greater than $10^{-5}-10^{-4}\:r_g$, nor smaller than $\sim 10^{-6}\:r_g$, 
but better modelling needs to be done. Nonetheless good estimates are
still possible without detailed knowledge of all the mechanisms in
accretion disks; this approach has the power to lead to a better
understanding of both kHz QPOs and other astrophysical phenomena.

\clearpage

\appendix
\section{Some notes on the numerical integration of SDEs} \label{sec:SDE}
\subsection{General remarks}

A generic system of SDEs can be written in the form
\begin{equation} \label{eq:sde1}
\xdb = \ab(t, \xb) + \Sigmab(t, \xb) \betab
\end{equation}
Here $\xb$, $\ab$ are $n$-dimensional column vectors, $\betab$ is a $m$-dimensional column vector
containing zero mean, unit variance, Gaussian white-noise processes and $\Sigmab$ is a 
$n \times m$ matrix. Typically, this
equation is written in the more rigorous form
\begin{equation} \label{eq:sde2}
d \xb = \ab(t, \xb) dt + \Sigmab(t, \xb) d \wb,
\end{equation}
with solution
\begin{equation} \label{eq:large}
x_t = x_{t_0} + \int_{t_0}^t \ab(s, x_s) d s + \int_{t_0}^t \Sigmab(s, \xb_s) d \wb_s.
\end{equation}
Here $\wb$ is a $m$-dimensional Wiener process.
The numerical integration of SDEs is quite a difficult problem. In fact, 
in the case of ordinary differential equations (ODEs) 
\begin{equation} \label{eq:ode}
d \xb = \ab(t, \xb) dt
\end{equation}
 numerical integration schemes are, either directly or indirectly, based on a Taylor expansion of the solution
\begin{equation}
\xb_t = \xb_{t_0} + \int_{t_0}^t \ab(s, \xb_s) d s.
\end{equation}
 Something similar holds also for SDEs. However, the stochastic counterpart of the deterministic Taylor 
expansion is rather complex. In order to understand this point without entering into overly technical arguments, it is instructive to compare the expansions relative to a one-dimensional autonomous version of Eq.~\eqref{eq:ode}
and of Eq.~\eqref{eq:sde2} with $m=1$.
In this case, for the ODE~\eqref{eq:ode} the first-order integral form of the Taylor formula 
in the interval
$[t_0, t]$ is
\begin{equation} \label{eq:sol1}
x_t = x_{t_0} + a(x_{t_0}) \int_{t_0}^t d s + R_2,
\end{equation} 
where $R_2$ is the remainder.
For the SDE~\eqref{eq:sde2} the corresponding expansion is
\begin{align} \label{eq:sol2}
x_t = x_{t_0} & + a(x_{t_0}) \int_{t_0}^t d s  + \sigma (x_{t_0}) \int_{t_0}^t d w_s \\
 & + \sigma'(x_{t_0}) \sigma (x_{t_0})
\int_{t_0}^t \int_{t_0}^s dw_z d w_s + \bar R,
\end{align} 
where the symbol ``$~{}'~$'' denotes differentiation with respect to $x$, and $\bar R$ is the remainder.
From Eq.~\eqref{eq:sol2} it is possible to see the presence of the additional terms
\begin{equation}
\int_{t_0}^t dw_s, \qquad \int_{t_0}^t \int_{t_0}^s dw_z dw_s.
\end{equation} 
When $n, m \neq 1$, it is possible to show that in the higher order expansions some quantities appear
as
\begin{equation}
I_{(j_1, j_2, \cdots, j_l)} = \int_{t_0}^t \int_{t_0}^{s_l} \cdots ~\int_{t_0}^{s_2} 
dw_{s_1}^{j_1} \cdots ~dw_{s_{l-1}}^{j_{l-1}} dw_{s_l}^{j_l},
\end{equation}
where $j_1, j_2, \cdots, j_l \in [0, 1, \ldots, m]$. Such quantities are termed
{\it multiple stochastic integrals}. The main problem in dealing with
them is that they cannot be computed exactly. Unfortunately, in its turn,
numerical approximation is also a difficult affair.

The consequence of this situation is that, even in the case of simple
systems, only integration schemes of very low order strong convergence \footnote{We shall say that
a discrete time approximation $x_{[k]}$ {\it converges strongly with order} $\gamma > 0$ at time $T$
if there exists a positive constant $C$, which does not depend on $\delta$, and a $\delta_0 > 0$ such that
$\epsilon(\delta) = {\rm E}(|x_T - x_{[T]} |) \leq C \delta^{\gamma}$ for each $\delta \in (0, \delta_0)$.}
can be used.
In fact, for the autonomous version of system~\eqref{eq:sde2} the most
commonly used technique is the Euler scheme
\begin{equation}
\xb_{[k+1]} = \xb_{[k]} +  \ab_{[k]} h_{[k]} + \Sigmab_{[k]} \Delta \wb_{[k]},  
\end{equation}
where $h_{[k]} = t_{[k+1]} - t_{[k]}$ is the integration time step at the time $t_{[k]}$, and the elements of 
the vector
\begin{equation} \label{eq:euler}
\Delta \wb_{[k]} = \int_{t_k}^{t_{k+1}} d \wb_t = \wb_{t_{[k+1]}} - \wb_{t_{[k]}} 
\end{equation}
are independent identically-distributed Gaussian random variables with mean equal to zero and 
variance equal to $h_{[k]}$.

\subsection{Small noise approximation}

If one takes into account that the order of strong convergence for the scheme~\eqref{eq:euler} is only $\gamma = 0.5$,
in contrast to $\gamma = 1$ for its deterministic counterpart, then it easy to understand why SDEs are not yet a 
standard tool in physical applications. 

In order to improve this situation, \citet{mil97} note that in many
problems the random fluctuations that
 affect a physical system are small. This means that the system~\eqref{eq:sde2}
can be written as 
\begin{equation} \label{eq:small}
d \xb = \ab(t, \xb) dt + \epsilon \Sigmab(t, \xb) d \wb,
\end{equation}
where $\epsilon$ is a small positive parameter. This is an important observation since, for small noise,
it is possible to construct special numerical methods that are more effective and easier to implement
than in the general case. In fact, the term of the expansion depends
not only on the time step $h$
but also on the parameter $\epsilon$. Typically, the mean-square global error of the schemes proposed
by \citet{mil97} is of order $\Om(h^p + \epsilon^k h^q)$ with $0 < q < p$. Although the strong order of these
methods is given by $q$, typically not a large number, they are able to reach high exactness because
of the factor $\epsilon^k$ at $h^q$. For example, the simple scheme
\begin{equation} \label{eq:rk0}
\xb_{[k+1]} = \xb_{[k]} + \frac{1}{6} (\Kb_1 + 2 \Kb_2 + 2 \Kb_3 + \Kb_4) + \epsilon \Sigmab \Delta \wb_{[k]} 
\end{equation}
where
\begin{align}
\Kb_1 & = h \ab(t_{[k]}, \xb_{[k]}), \label{eq:rk1} \\
\Kb_2 & = h \ab(t_{[k]} + h/2, \xb_{[k]} + \Kb_1 /2), \label{eq:rk2} \\
\Kb_3 & = h \ab(t_{[k]} + h/2, \xb_{[k]} + \Kb_2 /2), \label{eq:rk3} \\
\Kb_4 & = h \ab(t_{[k+1]}, \xb_{[k]} + \Kb_3), \label{eq:rk4}
\end{align}
is of order $\Om(h^4 + \epsilon h + \epsilon^2 h^{1/2})$. In other words, the order of strong convergence
is $0.5$, as for the Euler scheme, but better results are to be expected because of the term $\epsilon^2$
that multiplies $h^{1/2}$.

\begin{acknowledgements}
We thank Marek Abramowicz for his suggestions and support. The  discussions 
with Omer Blaes and Axel Brandenburg made this work possible.
 P.R. acknowledges Marco Ajello and Anna Watts for their  help and comments
and Sir Franciszek
 Oborski for the unique hospitality in his Castle during the Wojnowice Workshop (2005).
\end{acknowledgements}

\end{document}